\documentclass[9pt,twocolumn,twoside]{opticajnl}
\journal{opticajournal} % use for journal or Optica Open submissions

\setboolean{shortarticle}{true}
\usepackage{lineno}
\usepackage{xcolor}
\usepackage{float}
\usepackage{placeins}
\usepackage{legacy-styles/opticajournal}
\title{Topological Polarimetry: Polarization Skyrmions and Vortices from Tissue Scattering}

\author[1,*,+]{Elena Vasilieva}
\author[2,+]{Ifra Arif}
\author[2]{Oleksii Sieryi}
\author[1]{Alexander Doronin}
\author[2]{Alexander Bykov}
\author[3,*]{Igor Meglinski}
\affil[1]{Victoria University of Wellington, School of Engineering and Computer Science,   Wellington, 6140, New Zealand}
\affil[2]{University of Oulu, Opto-Electronics and Measurement Techniques,  Oulu, 90014, Finland}
\affil[3]{Aston University, Aston Institute of Photonic Technologies, Birmingham, B15 2TT, UK}
\affil[*]{Correspondence: elena.vasilieva@vuw.ac.nz\,\,\,\,and/or\,\,\,\,i.meglinski@aston.ac.uk}
\affil[+]{these authors contributed equally to this work}

\begin{abstract}
The Stokes-Mueller description of tissue polarimetry is conventionally interpreted through local observables such as birefringence, depolarization, and helicity preservation. We show that tissue structure generates topology in the polarization field itself. Polarization-resolved measurements of unstained human breast tissue reveal polarization vortices and, where the tissue architecture provides coupled azimuthal and radial variation, Néel-type polarization skyrmions. These structures are quantified by the vortex winding number and skyrmion charge, integer-valued topological invariants that vanish for homogeneous media and therefore provide zero-background readouts of tissue heterogeneity. Malignant ductal carcinoma exhibits non-trivial polarization topology, whereas adjacent healthy tissue remains topologically trivial. The results establish topological invariants of the polarization field as physically interpretable observables of tissue organization and motivate topological polarimetry as a field-based approach to tissue characterization.
\end{abstract}

\setboolean{displaycopyright}{false} % 

\begin{document}

\maketitle

%\section{Introduction}

Polarization imaging is traditionally interpreted through local observables. The Stokes-Mueller formalism describes how a medium transforms an incident polarization state into an output state~\cite{azzam2016}, and tissue polarimetry therefore reports birefringence, diattenuation, depolarization, and helicity preservation on a pixel-by-pixel basis~\cite{chao2021}. The pixels are treated independently, and the structure of the field they collectively form is not among the measured quantities. 

The spatial organization of the polarization field is set by the scattering medium~\cite{meglinski2025hidden}. A homogeneous, isotropic medium illuminated by a uniform circularly polarized laser light returns a statistically uniform back-scattered polarization field, however, strongly it depolarizes the incident state. Any spatial variation in the back-scattered field therefore originates from heterogeneity in the medium: a spatially organized scattering medium writes its organization into the geometry of the polarization field.

This geometry arises from polarization memory~\cite{mackintosh1989memory,xu2005memory}. Multiple scattering does not fully destroy circular polarization; it decomposes the back-scattered field into helicity-preserving and helicity-flipped components whose superposition defines the local polarization state~\cite{macdonald2015polydisperse}. Spatial variation in the balance of these two components writes a polarization texture onto the Poincar\'e sphere. The back-scattered field carries no genuinely unpolarized fraction: it separates into two fully polarized components of opposite helicity, so the polarization texture is built from fully defined states at every pixel rather than from a depolarized average~\cite{lopushenko2023depol}. This texture is topological: the polarization field carries vortices and skyrmions that are properties of its spatial organization, not of any individual pixel~\cite{shen2024review,flossmann2008speckle}.

These structures are quantified by two topological invariants. The polarization vortex winding number $\ell_C$~\cite{dennis2002singularities} is the circulation of the polarization azimuth around a closed contour, divided by $2\pi$: an integer counting the polarization singularities ($C$-points, $L$-lines) the contour encloses. The skyrmion charge $\mathcal{N}$~\cite{shen2024review} is the area integral of the Berry curvature of the polarization unit vector over the imaged region, divided by $4\pi$: an integer counting how many times the field wraps the Poincar\'e sphere. Because the polarization state is defined at every pixel, both invariants are computed on a genuine field rather than a depolarized average. They vanish identically for uniform illumination of a homogeneous medium, irrespective of the absolute output state, and being integer-valued they cannot be driven to a spurious non-zero value by measurement noise: they are zero-background readouts of the medium's heterogeneity, protected against calibration drift and illumination variation by their integer character.

\begin{figure*}[!h]
\centering
\includegraphics[width=0.85\textwidth]{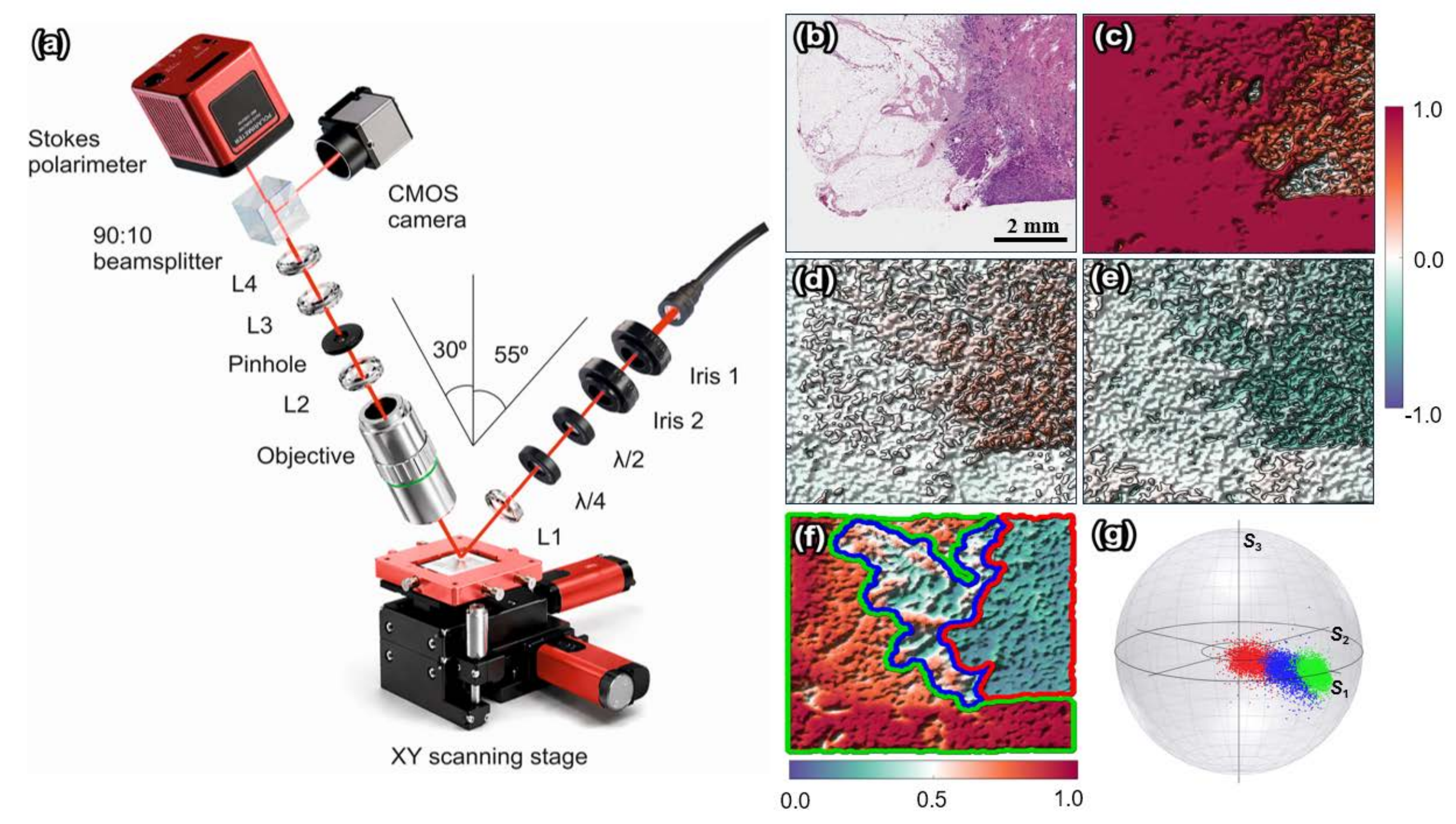}
\caption{\textbf{Polarimetric characterization of a FFPE tissue sample.} (a) Schematic representation of the tissue measurement system. Light from a supercontinuum laser (Leukos Ltd.) is filtered by an acousto-optic tunable filter to select a wavelength of $640~nm$, converted into right-circularly polarized light by a quarter-wave plate, and focused onto the tissue using a $45~mm$ lens. Back-scattered light is collected by a $100\times$ objective in confocal geometry through a $100~\mu m$ pinhole and analyzed by a Stokes polarimeter (Thorlabs Ltd.). CMOS camera coupled through a beam-splitter is used for focus alignment. The sample is raster-scanned on an $XY$ translation stage. (b) Representative histological image of the tissue section, containing invasive ductal carcinoma and adjacent healthy stroma, used for structural correlation with the polarization measurements. (c–e) Maps of the normalized Stokes parameters $S_1, S_2, S_3$; (f) Spatial DOP distribution, the fraction of light that remains polarized after interaction with the tissue; the outline demarcates the carcinoma lesion against the healthy background. (g) Poincaré sphere showing the per-pixel polarization states for the region demarcated in (f), colour-coded by tissue type (green: adipose; blue: fibrosis; red: carcinoma), with the adipose and fibrosis states clustered tightly and the carcinoma states spread across a broader region of the sphere. }
\label{fig:setup}
\end{figure*}

Here, we show that biological tissue generates such topology. Under uniform circularly polarized illumination, heterogeneous tissue produces polarization fields carrying non-trivial vortex winding and, where the tissue architecture supplies coupled azimuthal and radial variation, N\'eel-type polarization skyrmions, while homogeneous regions remain topologically trivial. We demonstrate this on unstained Formalin-fixed, paraffin-embedded (FFPE) human breast tissue containing ductal carcinoma, where the lesion generates non-trivial polarization topology and adjacent healthy tissue does not, establishing tissue scattering as a mechanism that generates topological states of light.
 
%\section{Materials and Methods}

Polarization-resolved measurements were performed using the custom imaging system shown in Fig.~\ref{fig:setup}(a). Right-circularly polarized illumination was focused onto the sample and the back-scattered Stokes vector $\mathbf{S}(x,y) = (S_0, S_1, S_2, S_3)^{T}$ was recorded at each scan position, where $S_0$ is the total intensity and $S_1, S_2, S_3$ are the horizontal-vertical, diagonal-antidiagonal, and circular polarization differences. The measurement geometry ($\theta_i = 55^\circ$, source-detector separation $\rho = 150~\mu m$) was selected to suppress specular reflection and probe superficial tissue microstructure in the quasi-ballistic back-scattering regime~\cite{LopushenkoJBO2024}, preferentially sampling Mie scatterers of radius $0.5 - 2~\mu m$, the size range of cell nuclei and nucleoli. FFPE human breast tissue sections (see Fig.~\ref{fig:setup}(b)) containing ductal carcinoma and adjacent healthy tissue were obtained from Biobank Borealis (Oulu) and annotated by a certified pathologist on the corresponding histological slide.

%Figure~\ref{fig:setup} maps the back-scattered polarized light pixel-by-pixel across the specimen: the degree of polarization (see Fig.~\ref{fig:setup}-b), the normalized Stokes parameters $S_1, S_2, S_3$ (see Fig.~\ref{fig:setup} c-e), and the distribution of polarization states on the Poincar\'e sphere (see Fig.~\ref{fig:setup}-g), shown alongside the corresponding histology (see Fig.~\ref{fig:setup}-f). These pixel-wise maps record how the lesion alters the back-scattered polarization, but do not separate organized structure from random depolarization.

Figure~\ref{fig:setup} maps the back-scattered polarized light pixel-by-pixel across the specimen. The degree of polarization (DOP, see Fig.~\ref{fig:setup}(f)) is reduced in the lesion relative to the healthy background, consistent with the increased depolarization reported for malignant tissue~\cite{ivanov2020,chao2021}. 
The normalized Stokes maps $S_1, S_2, S_3$ (see Fig.~\ref{fig:setup}(c-e)) show that healthy regions are dominated by a single linear-polarization axis while malignant regions develop components along the diagonal and circular axes, and on the Poincar\'e sphere (see Fig.~\ref{fig:setup}(g)) the healthy states cluster tightly while the malignant states spread across a broader region, correlated with the histology (see Fig.~\ref{fig:setup}(b)). The DOP, $S_1$, $S_2$ and $S_3$ maps in Fig.~\ref{fig:setup}(c–f) and the Poincaré-sphere distribution in Fig.~\ref{fig:setup}(g) are pixel-local descriptors: each value is computed from the four Stokes parameters of that pixel alone, so these representations report the per-pixel polarization state without resolving how the polarization field is organized in space. These pixel-wise maps therefore confirm that the lesion alters the back-scattered polarization but cannot separate organized structure from random depolarization.

\begin{figure*}[t]
\centering
\includegraphics[width=0.85\textwidth]{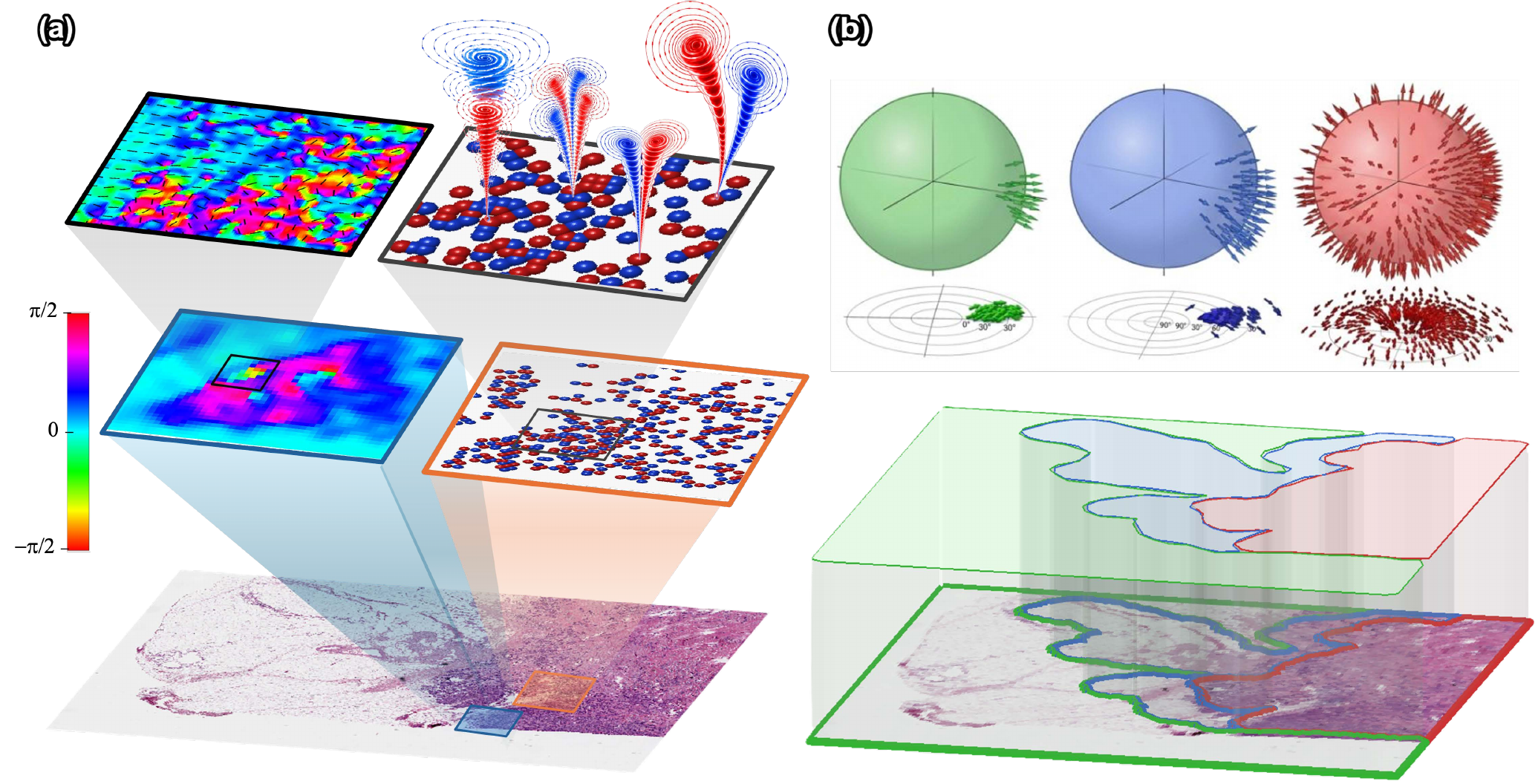}
\caption{\textbf{Topology of the back-scattered polarization field in healthy and malignant tissue.} \textbf{(a)} Exploded co-registration stack: H\&E histology of the imaged region (bottom, ROI boxed); local vortex-winding density $\rho_\ell$ (middle); and polarization azimuth field $\Psi(x,y)$ (top, hue = local ellipse orientation; black bars = director field), shown with two magnified insets. Magnified insets contrast healthy stroma with the IDC lesion, where singularities appear as red and blue phase vortices carrying winding $\ell_C = +1$ (red) and $-1$ blue. (b) Poincaré-sphere polarization texture for the three tissue types: carcinoma (red): $|N| = 1 - 4$ (non-zero integer across the fragments examined), against $|N| \rightarrow 0$ and $\ell_C \rightarrow 0$ for healthy stroma: adipose (green) \& fibrosis (blue). 
%Poincaré-sphere polarization textures of carcinoma (red) and healthy stroma: adipose (green) and fibrosis (blue). Carcinoma exhibits a non-zero integer topological charge ($|N| = 1 - 4$ across the examined fragments), whereas healthy tissue yields $|N| \rightarrow 0$ and $\ell_C \rightarrow 0$.
%adipose and fibrosis cluster near a single state, carcinoma covers a two-dimensional patch. %The plane above shows the local skyrmion density $\rho_{sk}$ over the lesion. Integrated over a malignant fragment $|N|$ is a non-zero integer (1 to 4 across the fragments examined), against $\ell_C \rightarrow 0$, $|N| \rightarrow 0$ for healthy stroma.
}
\label{fig:Skyrm}
\end{figure*}

The topological observables are constructed from the spatially resolved Stokes maps. The polarization azimuth~\cite{azzam2016}:
\begin{equation}
\Psi(x,y) = \tfrac{1}{2}\arctan_2(S_2, S_1)
\end{equation}
encodes the orientation of the major axis of the polarization ellipse and forms a scalar angle field on the imaged region. The line integral of its gradient around a closed contour $\mathcal{C}$ defines the polarization vortex winding number~\cite{flossmann2008speckle,dennis2002singularities}:
\begin{equation}
\ell_C = \frac{1}{2\pi}\oint_{\mathcal{C}} \nabla\Psi(x,y) \cdot d\mathbf{r},
\end{equation}
an integer counting the polarization singularities ($C$-points, $L$-lines) the contour encloses, identically zero for a region of uniform polarization orientation.

The reduced Stokes vector
\begin{equation}
\hat{\mathbf{n}}(x,y) = (S_1, S_2, S_3)/\sqrt{S_1^2 + S_2^2 + S_3^2}
\end{equation}
is a unit vector on the Poincar\'e sphere whose spatial distribution carries the complementary skyrmion charge~\cite{shen2024review},
\begin{equation}
\mathcal{N} = \frac{1}{4\pi}\iint_\Sigma \hat{\mathbf{n}} \cdot
\left(\frac{\partial\hat{\mathbf{n}}}{\partial x} \times
\frac{\partial\hat{\mathbf{n}}}{\partial y}\right) dx\,dy,
\end{equation}
an integer counting how many times the field wraps the Poincar\'e sphere. The two invariants encode distinct content: $\ell_C$ depends only on the azimuthal coordinate of $\hat{\mathbf{n}}$ and is non-zero for any in-plane polarization singularity, whereas $\mathcal{N}$ depends on both polar and azimuthal coordinates and is non-zero only when the texture covers the sphere as a two-dimensional patch rather than tracing a one-dimensional loop. Spatially organized variation of the local birefringent axis around a lesion generates vortex winding; additional radial variation of the helicity statistics, which connects the lesion centre to its periphery through different latitudes of the sphere, generates a N\'eel-type skyrmion texture.

Both invariants are evaluated on discrete maps by finite differences: $\ell_C$ by accumulating wrapped azimuth differences around each $2\times2$ cell, $\mathcal{N}$ by central differences of $\hat{\mathbf{n}}$. The local densities $\rho_\ell$ and $\rho_{\mathrm{sk}}$ are the corresponding per-pixel integrands, and their integrals over any region of interest, $\ell_C = \iint \rho_\ell\,dx\,dy$ and $\mathcal{N} = \iint \rho_{\mathrm{sk}}\,dx\,dy$, are integer-valued and instrument-independent.

%\section{Results and Discussion}

We demonstrate the topological observables on a representative FFPE human breast tissue section containing invasive ductal carcinoma (IDC), an early-stage malignancy where conventional histological contrast is subtle and polarization sensitivity is therefore most valuable.

The polarization azimuth~\cite{azzam2016}: $\Psi(x,y)=\tfrac{1}{2}\arctan_2(S_2,S_1)$, computed from the high-resolution Stokes maps, behaves qualitatively differently in healthy and malignant tissue (Fig.~\ref{fig:Skyrm}-a). Healthy stroma shows a slowly varying azimuth aligned with the local collagen orientation, giving near-zero loop integrals of $\nabla\Psi$. Cancerous regions contain localized polarization singularities at which $\psi$ is undefined and around which it winds by $\pm\pi$ or $\pm2\pi$~\cite{dennis2002singularities}. The local winding density $\rho_\ell$ resolves these as point-like defects spatially correlated with the duct boundary and disrupted collagen architecture. Integrated around a contour enclosing the lesion, the winding number is $\ell_C \approx \pm 1$, against $\ell_C \to 0$ for any comparable contour in healthy stroma. This non-zero integer is a direct signature of tissue-generated singularities and depends only on organized variation of the local birefringent axis around the lesion.

The skyrmion charge is a more demanding indicator of tissue microstructure (Fig.~\ref{fig:Skyrm}-b). The local density $\rho_{\mathrm{sk}}$ is approximately zero in healthy regions, where the texture is nearly uniform on the Poincar\'e sphere, and forms a coherent non-zero pattern in malignant regions where the heterogeneity generates dual-axis variation of $\hat{\mathbf{n}}$~\cite{shen2024review}. The integrated topological charge is an additive integer that counts spatially separated skyrmionic textures. In malignant tissue $|N|$ ranges from 1 to 4, increasing with lesion extent and heterogeneity; within a given fragment it remains stable and robust to illumination power, and it vanishes in healthy tissue ($|N| \rightarrow 0$). Where the charge is non-zero the texture is of N\'eel type, with the azimuth of $\hat{\mathbf{n}}$ locked to the spatial polar angle around the lesion and the polar coordinate varying from centre to periphery.

The two outcomes correspond to distinct tissue geometries. A vortex without an accompanying skyrmion ($\ell_C \neq 0$, $\mathcal{N}=0$) arises when the tissue modulates only the orientation of the local birefringent axis: $\hat{\mathbf{n}}$ circulates in azimuth around a singularity but stays at a fixed latitude of the Poincar\'e sphere, as produced by disrupted collagen with no change in helicity preservation. A skyrmion ($\mathcal{N}\neq 0$) requires the tissue to vary, in addition, the degree of helicity survival across the region, so that $\hat{\mathbf{n}}$ sweeps in latitude from the lesion centre to its periphery while winding in azimuth, covering a two-dimensional patch of the sphere. This radial degree of freedom is supplied by the gradient in scatterer size and refractive-index contrast that accompanies malignant nuclear enlargement, shifting local scattering from the Rayleigh toward the Mie regime~\cite{Perelman_PRL} and altering circular-polarization survival~\cite{macdonald2015polydisperse}. Healthy stroma, lacking both the orientational disruption and the helicity gradient, generates neither; the IDC lesion exhibits both. In geometric terms the two invariants read independent coordinates of $\hat{n}$ on the Poincaré sphere: $\ell_C$ is fixed by the azimuth alone, whereas a non-zero $N$ requires a second, non-collinear gradient in the polar angle set by helicity survival~\cite{macdonald2015polydisperse}, so a vortex reports orientational disorder while a skyrmion reports its coincidence with the nuclear-scale scattering gradient of malignancy~\cite{Perelman_PRL}.

The skyrmion charge is therefore the more specific marker: it is non-zero only where orientational disorder and a helicity gradient coincide, a condition met by the coupled architectural and nuclear-scale changes of malignancy but not by orientational disruption alone. Both invariants vary between healthy and malignant tissue by far more than the measurement uncertainty, and as integer-valued quantities they are immune to the calibration drift and illumination variation that limit scalar polarimetric parameters. Tissue heterogeneity is the sole source of their non-zero values. The topological invariants reported here are not formal constructs imposed on partially polarized data: they are quantized properties of the hidden coherent structure that multiply-scattered circularly polarized light retains. The back-scattered field decomposes rigorously into two fully polarized components of opposite helicity with stable phase couplings~\cite{lopushenko2023depol,meglinski2025hidden}, so the azimuth field $\Psi(x,y)$ and the reduced Stokes vector $\hat{\mathbf{n}}(x,y)$ read out the spatial phase structure of that coherence rather than statistical averages over a depolarized ensemble. The invariants $\ell_C$ and $N$ are therefore integer-valued topological charges of the hidden coherence itself, and their non-vanishing in malignant tissue is direct evidence that biological heterogeneity organizes the surviving phase coherence of the scattered field into topologically protected textures. Tissue scattering does not destroy polarization coherence; it organizes it into quantized topological textures. The topological invariants are therefore complementary to the pixel-local descriptors of Fig.~\ref{fig:setup}, and not a replacement for them: DOP and the normalized Stokes components report the polarization state at each pixel, whereas $\ell_C$ and $N$ report how those states are organized across pixels. These two readouts probe orthogonal content of the same dataset and, in combination, distinguish organized from random heterogeneity.

That a structureless input field can acquire non-trivial polarization topology through a deterministic physical process is established for coherent free-space propagation, where a spin-balanced beam develops full Poincar\'e-sphere coverage governed by a single topological index~\cite{mkhumbuza2026}. Here the generator is not propagation but scattering: the spatial heterogeneity of the tissue plays the role of the topological control parameter, mapping disordered birefringent and scattering organization onto the coverage of the sphere.

%\section{Conclusion}
In conclusion, we demonstrate that biological tissue can generate non-trivial topological states in the polarization field of back-scattered light. By quantifying the field through the polarization-vortex winding number and skyrmion charge, we identified polarization vortices and Néel-type polarization skyrmions in unstained human breast tissue and showed that malignant ductal carcinoma produces topologically non-trivial polarization textures, whereas adjacent healthy tissue remains topologically trivial.

These topological invariants constitute a new class of polarimetric observables. Unlike conventional scalar metrics, which quantify the polarization state at individual pixels, the winding number and skyrmion charge quantify the spatial organization of the polarization field itself. As integer-valued quantities, they possess an intrinsic zero background, are insensitive to absolute intensity, and exhibit robustness against calibration drift and illumination fluctuations.

More fundamentally, the results establish tissue heterogeneity as a physical mechanism for generating topological states of light. Scattering is therefore not merely a source of depolarization but also a mechanism through which biological structure is encoded into the topology of the polarization field. This shifts polarimetry from the analysis of local polarization states to the characterization of field topology and introduces topological polarimetry as a physically interpretable framework for probing tissue organization across multiple spatial scales. The ability to quantify organized and disorganized heterogeneity through topological invariants opens new opportunities for label-free tissue characterization and optical diagnostics.

%In conclusion, we have shown that biological tissue generates topology in the polarization field of back-scattered light. By quantifying this topology through the vortex winding number and skyrmion charge, we identified polarization vortices and Néel-type polarization skyrmions in unstained human breast tissue and demonstrated that malignant regions exhibit non-trivial topological textures whereas adjacent healthy tissue remains topologically trivial. These topological invariants provide zero-background observables of tissue heterogeneity that are intrinsically robust to calibration drift and illumination fluctuations. Topological polarimetry is complementary to scalar polarimetric markers rather than a replacement: scalar observables report the local polarization state at each pixel, the topological invariants report the spatial organization of the polarization field across pixels, and the combination adds a diagnostic axis, namely organized versus unorganized heterogeneity, that neither delivers alone. More broadly, the results establish tissue heterogeneity as a source of topological states of light and extend conventional polarimetry from the analysis of local polarization states to the topology of polarization fields, opening a route toward topological polarimetry as a physically interpretable framework for tissue characterization.

\begin{backmatter}
\bmsection{Funding} 
This article is based upon work from the Horizon Europe EIC Pathfinder Open Research and Innovation Programme, OPTIPATH project, Grant Agreement No. 101185769.

\bmsection{Acknowledgment} 
Authors acknowledge partial support of the European Union under the Marie Skłodowska-Curie Action I4WORLD 10108128. 
The authors thank Biobank Borealis (Finland) and Dr. Juha Näpänkangas (Oulu University Hospital, Finland) for providing tissue samples and performing their histological characterization.

\bmsection{Disclosures} The authors declare no conflicts of interest.

\bmsection{Data availability} 
Data underlying the results presented in this paper are not publicly available at this time but may be obtained from the authors upon reasonable request.
\end{backmatter}

%\section{References}
\bibliography{sampleO}

@article{azzam2016,
  author  = {Azzam, R. M. A.},
  title   = {Stokes-vector and {Mueller}-matrix polarimetry [Invited]},
  journal = {J. Opt. Soc. Am. A},
  volume  = {33},
  pages   = {1396--1408},
  year    = {2016}
}

@article{chao2021,
  author  = {Chao, H. and He, H. and Chang, J. and Chen, B. and Ma, H. and Booth, M. J.},
  title   = {Polarisation optics for biomedical and clinical applications: a review},
  journal = {Light Sci. Appl.},
  volume  = {10},
  pages   = {194},
  year    = {2021}
}

@article{meglinski2025hidden,
  author  = {Meglinski, I. V. and Tuchin, V. V.},
  title   = {Hidden coherent structure and phase correlations of unpolarized light in multiple scattering},
  journal = {Dokl. Ross. Akad. Nauk. Fiz. Tekh. Nauki},
  volume  = {524},
  pages   = {23--32},
  year    = {2025}
}

@article{LopushenkoJBO2024,
author = {Ivan Lopushenko and Oleksii Sieryi and Alexander Bykov and Igor Meglinski},
title = {{Exploring the evolution of circular polarized light backscattered from turbid tissue-like disperse medium utilizing generalized Monte Carlo modeling approach with a combined use of Jones and Stokes-Mueller formalisms}},
volume = {29},
journal = {J. Biomed. Opt.},
number = {5},
pages = {052913},
year = {2024}
}

@article{shen2024review,
  author  = {Shen, Y. and Mart{\'i}nez, E. C. and Rosales-Guzm{\'a}n, C. and others},
  title   = {Optical skyrmions and other topological quasiparticles of light},
  journal = {Nat. Photonics},
  volume  = {18},
  pages   = {15--25},
  year    = {2024}
}

@article{flossmann2008speckle,
  author  = {Flossmann, F. and O'Holleran, K. and Dennis, M. R. and Padgett, M. J.},
  title   = {Polarization singularities in {2D} and {3D} speckle fields},
  journal = {Phys. Rev. Lett.},
  volume  = {100},
  pages   = {203902},
  year    = {2008}
}

@article{dennis2002singularities,
  author  = {Dennis, M. R.},
  title   = {Polarization singularities in paraxial vector fields: morphology and statistics},
  journal = {Opt. Commun.},
  volume  = {213},
  pages   = {201--221},
  year    = {2002}
}

@article{mkhumbuza2026,
  author  = {Light Mkhumbuza and Pedro Ornelas and Angela Dudley and Isaac Nape and Kayn A. Forbes}, 
  title   = {Topological control of chirality and spin with structured light},
  journal = {Light Sci. Appl.},
  volume  = {15},
  pages   = {214},
  year    = {2026}
}

@article{ivanov2020,
author = {Ivanov, Deyan and Dremin, Viktor and Bykov, Alexander and Borisova, Ekaterina and Genova, Tsanislava and Popov, Alexey and Ossikovski, Razvigor and Novikova, Tatiana and Meglinski, Igor},
title = {{Colon cancer detection by using Poincaré sphere and 2D polarimetric mapping of \textit{ex vivo} colon samples}},
journal = {J. Biophotonics},
volume = {13},
number = {8},
pages = {e202000082},
year = {2020}
}

@article{Perelman_PRL,
  title = {Observation of Periodic Fine Structure in Reflectance from Biological Tissue: A New Technique for Measuring Nuclear Size Distribution},
  author = {Perelman, L. T. and Backman, V. and Wallace, M. and Zonios, G. and Manoharan, R. and Nusrat, A. and Shields, S. and Seiler, M. and Lima, C. and Hamano, T. and Itzkan, I. and Van Dam, J. and Crawford, J. M. and Feld, M. S.},
  journal = {Phys. Rev. Lett.},
  volume = {80},
  issue = {3},
  pages = {627--630},
  year = {1998}
}

@article{lopushenko2023depol,
  author  = {Lopushenko, I. and Bykov, A. and Meglinski, I.},
  title   = {Depolarization composition of backscattered circularly polarized light},
  journal = {Phys. Rev. A},
  volume  = {108},
  pages   = {L041502},
  year    = {2023}
}

@article{mackintosh1989memory,
  author  = {MacKintosh, F. C. and Zhu, J. X. and Pine, D. J. and Weitz, D. A.},
  title   = {Polarization memory of multiply scattered light},
  journal = {Phys. Rev. B},
  volume  = {40},
  pages   = {9342},
  year    = {1989}
}

@article{xu2005memory,
  author  = {Xu, M. and Alfano, R. R.},
  title   = {Circular polarization memory of light},
  journal = {Phys. Rev. E},
  volume  = {72},
  pages   = {065601(R)},
  year    = {2005}
}

@article{macdonald2015polydisperse,
  author  = {Macdonald, C. M. and Jacques, S. L. and Meglinski, I. V.},
  title   = {Circular polarization memory in polydisperse scattering media},
  journal = {Phys. Rev. E},
  volume  = {91},
  pages   = {033204},
  year    = {2015}
}

%\bibliographyfullrefs{sampleO}

\end{document}